# Rayleigh-Sommerfeld Fraunhofer Diffraction

Robert L. Lucke
Code 7231
Naval Research Laboratory
Washington DC 20375
robert.lucke@nrl.navy.mil
202-767-2749

Abstract

Treatments of Fraunhofer diffraction normally approximate the inclination factor in the diffraction integral by 1, but this is not necessary. In this paper, the Rayleigh-Sommerfeld diffraction integral is used and a complete description of Fraunhofer diffraction is given, with the minimum possible number of approximations made. A focused wave is examined before, at, and after its focal point to elucidate the exact role of the Fourier transform in describing Fraunhofer diffraction and to derive the Guoy phase shift.

## 1. Introduction

Fraunhofer diffraction is fundamental to any introductory college-level optics course and should be taught with as much precision and as few approximations as possible. The archetypical treatment of Fraunhofer diffraction is that of Born and Wolf[1], which starts with the Fresnel-Kirchhoff (FK) diffraction integral and must therefore approximate the inclination factor (also called the obliquity factor) by 1 before proceeding to the Fraunhofer analysis. Without this approximation, analytic progress cannot be made. Another popular treatment is that of Goodman[2], which makes additional approximations (see below). But if we begin with the Rayleigh-Sommerfeld (RS) diffraction integral, it is not necessary to approximate the inclination factor by 1. In a recent paper[3], I showed the superiority of RS over FK for providing an accurate description of Poisson's spot. (The two integrals use different inclination factors, but are otherwise the same.) I now use the RS diffraction integral to give the most accurate possible description of Fraunhofer diffraction. Once this is done, the approximations of Born and Wolf and of Goodman can be made, as needed or desired, with a full understanding of their accuracy and consequences.

With reference to Figure 1 (next page), Fraunhofer diffraction describes both (1) the far-field diffraction pattern of a small source and (2) the focal plane illumination pattern of a converging wave. These may be referred to as the forward and inverse Fraunhofer problems, respectively. After setting the inclination factor to 1, Born and Wolf do the forward problem and show that the Fourier transform relation between the source wave front and the calculated wave front is from Cartesian coordinates to direction cosines, but they derive the inverse problem from the forward problem in a way that obscures the fact that the Fourier transform relation in the inverse problem is from direction cosines to Cartesian coordinates. Goodman also sets the inclination factor to 1, then uses an approximation that causes the Fourier transform relation to be between two sets of Cartesian coordinates. The Born and Wolf treatment is also subject to the mild criticism that it doesn't keep track of the overall phase factor of $-i$ outside the Fraunhofer integral.

This paper keeps the inclination factor in the calculation and does the forward and inverse problems separately to obtain the more accurate results that, in case (1), the diffraction pattern is proportional to the Fourier transform of the source wave front multiplied by the inclination factor, while, in case (2), the source wave front is divided by the inclination factor before the Fourier transform is applied. Also, this paper combines the two problems to show the relation of an after-focus wave front to a before-focus wave front, a relation that includes the Guoy phase shift. By this means, the Guoy phase shift can be derived from Fraunhofer diffraction.

Sec. 2 treats the basic Fraunhofer diffraction problem of a wave diverging from a small source. Sec. 3 treats the inverse problem of a wave converging on a focal point and exhibits the converging wave that results in the traditional Airy-disc diffraction pattern (it has a $\cos\theta$ amplitude dependence), and Sec. 4 combines the two to show what happens when a wave passes through a focal point.

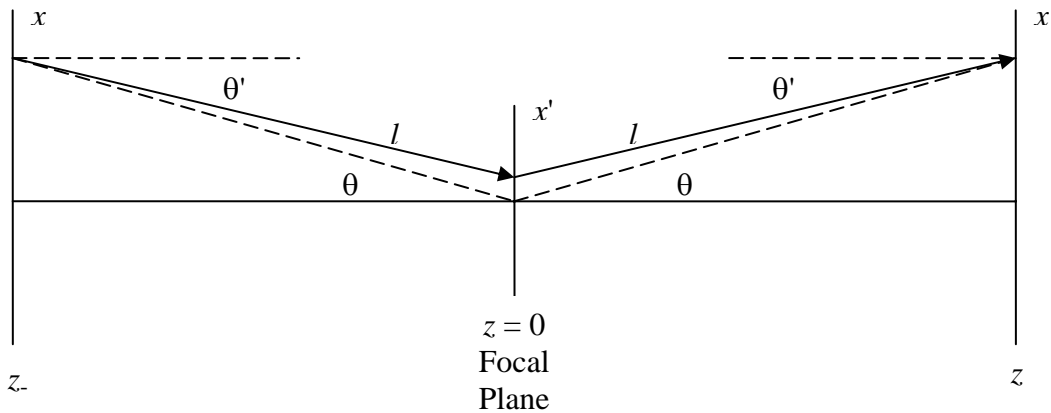

Figure 1. A wave converging from the left is brought to a focus in the $z = 0$ plane, then diverges to the right. $z$ and $z_-$ are sufficiently far from the focal plane to assure that the conditions of Fraunhofer diffraction are met for both processes. The right side of the figure illustrates waves diverging from a small source, as described in Sec. 2, the left side shows converging waves, as described in Sec. 3. Both sides are used in Sec. 4. The planes at $z$ and $z_-$ both use $(x, y)$ coordinates (the $+y$ axis points out of the page), a fact that will not confuse the alert reader. On the right side of the figure, angle $\theta$ is defined by $\tan\theta = (x^2 + y^2)^{1/2}/z$, on the left by $\tan\theta = (x^2 + y^2)^{1/2}/(-z_-)$ so that $0 \leq \theta \leq \pi/2$ always; again this will cause no confusion. The vectors of length $l$ point from an arbitrary point in the source wave front to an arbitrary point in the calculated wave front. In both cases the angle $\theta'$ is the angle between this vector and the $+z$ direction.

## 2. The Forward Fraunhofer Problem: A Wave Diverging from a Small Source

We begin with the standard Fraunhofer problem of calculating the far-field diffraction pattern of a light wave diverging from a small source. This is illustrated on the right side of Figure 1 for light diverging from a focal plane, but the derivation is valid for any small diffractive source, such as an aperture illuminated by a plane wave. The RS diffraction integral[2,3] for this case is

$$U(x, y, z) = \frac{-i}{\lambda} \int_{x',y'} U(x', y', 0) \cos\theta' \frac{\exp(ikl)}{l} dx'dy'$$
$$= \frac{-i}{\lambda} \int_{x',y'} U(x', y', 0) \frac{z}{l^2} \exp(ikl) dx'dy' \quad,$$
(1)

where $k = 2\pi/\lambda$, $l^2 = (x - x')^2 + (y - y')^2 + z^2$, and the inclination factor is $\cos\theta' = z/l$. Now, using $r^2 = x^2 + y^2 + z^2$,

$$\begin{aligned}
l &= \left[ x^2 - 2xx' + x'^2 + y^2 - 2yy' + y'^2 + z^2 \right]^{1/2} \\
&= \left[ r^2 - 2(xx' + yy') + x'^2 + y'^2 \right]^{1/2} \\
&= r - \frac{xx' + yy'}{r} + \frac{x'^2 + y'^2}{2r} + O\left( \frac{x^2 x'^2}{r^3}, \frac{y^2 y'^2}{r^3} \right) \\
&\approx r - \frac{xx' + yy'}{r} \quad.
\end{aligned}$$
(2)

In the argument of the exponential function in Eq. (1), any error in approximating $l$ must be small compared to $\lambda$ in order to give an accurate result for the integral. To see this, we need only observe that $\exp[ik(l + \delta l)] = \exp(ikl) \times \exp(ik\delta l)$ and that $\exp(ik\delta l) \approx 1 + i2\pi\delta l/\lambda$ for small $\delta l$. Defining $a^2$ as the maximum value of $x'^2 + y'^2$, it follows that $a^2/2r$ is the maximum possible value of the lowest-order neglected term in Eq. (2). Therefore, the approximation preserves phase as long as $\delta l = a^2/2r \ll \lambda/2\pi$ or

$$r \gg \frac{\pi a^2}{\lambda} \quad.$$
(3)

Outside the exponential, $l$ appears only as the multiplicative factor $l^{-2}$. To illustrate the effect of an approximation in this factor, we set $y = 0$, use $x/r < 1$, $x' \leq a$, and $a^2/2r \ll \lambda/2\pi$ to see from Eqs. (2) and (3) that

$$1 - \frac{l}{r} \approx \frac{x}{r}\frac{x'}{r} < \frac{a}{r} < \frac{\sqrt{\lambda r/\pi}}{r} = \sqrt{\frac{\lambda}{\pi r}} \quad,$$
(4)

which shows that the ratio $l/r$ can be made as close to 1 as desired, *i.e.*, we may set $l \approx r$, if $r$ is big enough. Specifically, if, say, $r > 10^4\lambda$, then $l$ differs from $r$ by less than 1%. For large $r$, therefore, only the first term in the approximation in Eq. (2) need be used when $l$ appears outside the exponential.

Using both terms in the last line of Eq. (2) in the exponent and only the first term outside it are the essential Fraunhofer approximations, and are the only ones made in this paper. Putting these approximations into Eq. (1) and noting that $z/r = \cos\theta$ gives

$$\begin{aligned}
U(x,y,z) &\approx \frac{-i}{\lambda} \iint_{x',y'} U(x',y',0) \frac{z}{r^2} \exp\left[ik\left(r - \frac{xx' + yy'}{r}\right)\right] dx'dy' \\
&= \frac{-i}{\lambda} \cos\theta \frac{\exp(ikr)}{r} \iint_{x',y'} U(x',y',0) \exp\left[-i2\pi\left(\frac{x}{\lambda r}x' + \frac{y}{\lambda r}y'\right)\right] dx'dy' \\
&= \frac{-i}{\lambda} A_0\left(\frac{x}{\lambda r}, \frac{y}{\lambda r}\right) \cos\theta \frac{\exp(ikr)}{r} \\
&= \frac{-i}{\lambda} A_0\left(\frac{\sin\theta\cos\phi}{\lambda}, \frac{\sin\theta\sin\phi}{\lambda}\right) \cos\theta \frac{\exp(ikr)}{r} ,
\end{aligned} \qquad (5)$$

where (using Goodman's notation) $A_0$ is the Fourier transform of $U(x', y', 0)$, $x/r = \sin\theta\cos\phi$ and $y/r = \sin\theta\sin\phi$ have been used, and the last form is given to emphasize that in the Fraunhofer regime, $A_0$ is a function only of direction, not of distance. In Eq. (5), most of the phase information about $U(x, y, z)$ is contained in the $\exp(ikr)$ factor, which describes a spherical phase front. But this phase front can be modified by $A_0$: in directions $(\theta, \phi)$ where $A_0$ is negative phase is reversed, and if $A_0$ is a complex function phase can change (usually slowly!) with direction. In either case, phase fronts are no longer strictly spherical.

The quantities $x/r$ and $y/r$ are direction cosines[1,2]. Observe that the Fourier transform in Eq. (5) is between the coordinate sets $(x', y')$ and $(x, y)/\lambda r$, *i.e.*, between the Cartesian coordinates of the source wave front and the direction cosines, with $\lambda$ acting as a scale factor for the latter: for the same source size, longer wavelengths diffract into a larger range of direction cosines, *i.e.*, into larger angles. Eq. (5) shows that the Fourier-transformed quantity is the source wave front, as expected, but that the amplitude of the resulting wave is proportional to this Fourier transform multiplied by $\cos\theta$, not just the Fourier transform itself.

The effect of the approximations made in other treatments of Fraunhofer diffraction can be seen by inspection of Eq. (5). First and most obvious, all other treatments known to this writer approximate the inclination factor by 1, which means that the $\cos\theta$ factor does not appear. (The interested student is invited to repeat the above calculation and that of the next section with the RS inclination factor replaced by the FK inclination factor.) To be sure, this improvement in accuracy is very small for the values of $\theta$ usually encountered in optical diffraction problems, but see Lucke[3], where Poisson's spot is presented as a counterexample.

Second, with reference to the second line of Eq. (5), Goodman[2], in his Eq. (4-13) on p. 61, uses the parabolic-wave approximation (so-called because it implies a parabolic rather than a spherical wave front), that is, he uses $r \approx z + (x^2 + y^2)/2z$ in the exponent outside the integral. This expansion neglects the next higher term, which is $-(x^2 + y^2)^2/8z^3$, and can lead to large errors in describing phase. To see this, we set $y = 0$, write $\theta \approx x/z$, and set the value of the neglected term to one-half wavelength: $x^4/8z^3 \approx z\theta^4/8 = \lambda/2$. Solving this equation for $z$ yields $z = 4\lambda/\theta^4$, which shows that when $\theta \neq 0$ there is a distance $z$ at which the phase error becomes large. When $\theta = 2° \approx 0.035$ rad, this happens for $z \approx 2.7\times10^6 \lambda$, between one and two meters for optical wavelengths. There are many problems in optics for which this phase error is not important, but those who use the parabolic-wave approximation should be aware of it.

Third, Goodman uses the approximation $1/r \approx 1/z$, both outside the integral and in the exponent inside the integral. This means that the $1/r$ dependence of amplitude on distance in Eq. (5) becomes $1/z$ dependence, which affects only amplitude and is quite accurate as long as $\theta$ is small. Since $x/z = \tan\theta\cos\phi$ and $y/z = \tan\theta\sin\phi$, it also means that $\tan\theta$ replaces $\sin\theta$ in the argument of the Fourier transform, which slightly changes the direction in which features of the diffraction pattern appear.

## 3. The Inverse Fraunhofer Problem: A Wave Converging to a Focus

With reference to the left side of Fig. 1, we start with a wave front with arbitrary angular dependence converging from the left on a focal point, and use Eq. (1) to find the amplitude in the focal plane. In this case, $\cos\theta' = (-z_-)/l$, and the integral is over $(x, y)$. The initial wave front is described by $U(x, y, z_-) = U_0 f(\theta, \phi)\exp(-ikr)/r$, where the minus sign in the exponent describes a wave traveling to the right (because, when $z < 0$, $r$ decreases with time rather than increasing). The RS diffraction integral is:

$$\begin{aligned} U(x', y', 0) &= \frac{-i}{\lambda} \iint_{x,y} U(x, y, z_-) \cos\theta' \frac{\exp(ikl)}{l} dxdy \\ &= \frac{-i}{\lambda} \iint_{x,y} U_0 f(\theta, \phi) \frac{\exp(-ikr)}{r} \cos\theta' \frac{\exp(ikl)}{l} dxdy \\ &\approx \frac{-i}{\lambda} U_0 \iint_{x,y} \frac{f(\theta, \phi)}{\cos\theta} \exp\left[-ik\frac{xx' + yy'}{r}\right] \frac{\cos^2\theta}{r^2} dxdy \quad, \end{aligned} \qquad (6)$$

where exactly the same approximations have been made in the last line of Eq. (6) as were made in Eqs. (2) and (5): the $(x'^2 + y'^2)/2r$ term in Eq. (2) makes a negligibly small contribution to phase if $z_-$ is big enough (the fact that $x'$ and $y'$ are no longer the variables of integration is irrelevant), and $l \approx r$ outside the exponent so that $\cos\theta' = (-z_-)/l \approx (-z_-)/r = \cos\theta$.

The integration is now over $(x, y)$, so it isn't obvious that Eq. (6) is a Fourier transform of the function $f(\theta, \phi)/\cos\theta$. In fact, it is not a Fourier transform over the variables $(x, y)$, but it *is* a Fourier transform over the direction cosines, that is, over the variables $(x/r, y/r)$. To show this, we make the substitutions $u = x/r$, $v = y/r$. Observe that $-1 \leq u, v \leq 1$ and that $u^2 + v^2 \leq 1$. Now $\partial u/\partial x = (r^2 - x^2)/r^3$, *etc.*, so

$$dudv = \begin{vmatrix} \frac{\partial u}{\partial x} & \frac{\partial u}{\partial y} \\ \frac{\partial v}{\partial x} & \frac{\partial v}{\partial y} \end{vmatrix} dxdy = \frac{1}{r^6} \begin{vmatrix} r^2 - x^2 & -xy \\ -xy & r^2 - y^2 \end{vmatrix} dxdy = \frac{z^2}{r^4} dxdy = \frac{\cos^2\theta}{r^2} dxdy \quad, \qquad (7)$$

where $z = r\cos\theta$ has been used, and Eq. (6) becomes

$$U(x', y', 0) = \frac{-i}{\lambda} U_0 \iint_{u^2+v^2 \leq 1} \frac{f(\theta, \phi)}{\cos\theta} \exp[-ik(ux'+vy')] du dv \quad . \tag{8}$$

Eq. (8) has the form of a Fourier transform, although the purist may want to extend the integration limits to $\pm \infty$. This is easily accomplished by observing that, since $u = x/r = \sin\theta\cos\phi$ and $v = y/r = \sin\theta\sin\phi$, it follows that $\sin\theta = (u^2 + v^2)^{1/2}$ and $\tan\phi = v/u$, and these equations can be inverted to give $\theta(u, v)$ and $\phi(u, v)$. We now define a new function $g(u, v) = f[\theta(u, v), \phi(u, v)]/\cos\theta(u, v)$ for $u^2 + v^2 \leq 1$, $g(u, v) = 0$ otherwise, to extend the integration limits to $\pm \infty$. Thus the assertion that the Fourier-transformed quantity is the source wave front divided by $\cos\theta$ is proved.

Eq. (5) is a Fourier transform from Cartesian coordinates to direction cosines, while Eq. (8) is from direction cosines to Cartesian coordinates. But one is not the inverse of the other: both are *forward* Fourier transforms. This point will be revisited in Sec. 4.

The interested reader may easily verify that if both the approximations $\cos\theta \approx 1$ and $1/r \approx 1/z$ are made in the last line of Eq. (6), then the substitutions $u = x/z$, $v = y/z$ give Eq. (8) without the $1/\cos\theta$ factor in the integrand. This is the approximation used by Goodman[2] for the inverse Fraunhofer problem [see his Eq. (5-45) with $\tilde{x}, \tilde{y}$ taken from Eq. (5-35), but note that these are Cartesian coordinates, not direction cosines]. Observe that if the converging wave is formed by an optical system with an *f*-number of 1, then the maximum value of $\theta$ is 26° and that $\cos 26° \approx 0.9$, which can be expected to make a noticeable, though not large, difference in the value of the integral in Eq. (8).

For the special case of a circularly symmetric wave front, Eq. (6) can be reduced to a one-dimensional integral. In this case, $f(\theta, \phi) = f(\theta)$ and the symmetry of the problem says that we can set $y' = 0$ and $x' = \rho'$, where $\rho'$ is the radial coordinate in the focal plane. We now use $dxdy = \rho d\phi d\rho$, where $\rho$ is the radial coordinate in the $(x, y)$ plane, $\rho = z\tan\theta$, $d\rho/d\theta = z/\cos^2\theta$, $x = r\sin\theta\cos\phi$, and $r = z/\cos\theta$ so that $\cos^2\theta dxdy/r^2 = \cos\theta\sin\theta d\theta d\phi = \sin\theta d(\sin\theta)d\phi$. Using the definition of the $J_0$ Bessel function, Eq. (6) becomes

$$\begin{aligned} U(\rho', 0) &= \frac{-i}{\lambda} U_0 \int_0^{2\pi} \int_0^1 \frac{f(\theta)}{\cos\theta} \exp(-ik\rho'\sin\theta\cos\phi) \sin\theta d(\sin\theta) d\phi \\ &= \frac{-i}{\lambda} U_0 2\pi \int_0^1 \frac{f(\theta)}{\cos\theta} J_0(k\rho' u) u du \quad , \end{aligned} \tag{9}$$

where the substitution $u = \sin\theta$ has been made. {With reference to the first equation on p. 248 of Bracewell[5], Eq. (9) can be seen to have the form of the 2D Fourier transform of $f(\theta)/\cos\theta$, a statement that will not be dwelt on here because it is shown in general in Eq. (8).}

A converging wave front with a convergence half-angle of $\theta_0$ and $\cos\theta$ angular dependence is described by $U(x, y, z_-) = U_0 \cos\theta R(\theta/\theta_0) \exp(-ikr)/r$, where $R$ is the rectangle function, $R(u) = 1$ for $|u| \leq 1$, $R(u) = 0$ for $|u| > 1$. Putting $f(\theta) = \cos\theta R(\theta/\theta_0)$ in Eq. (9), we find, substituting $v = k\rho' u$ and using the Bessel function relation given in Eq. (11) on p. 395 of Born & Wolf[1], that

$$U(\rho', 0) = -ikU_0 \int_0^{\sin\theta_0} J_0(k\rho' u) u \, du = -ikU_0 \frac{1}{(k\rho')^2} \int_0^{k\rho'\sin\theta_0} v J_0(v) \, dv$$

$$= -ikU_0 \frac{1}{(k\rho')^2} \left[ v J_1(v) \right]_0^{k\rho'\sin\theta_0} = -iU_0 \sin\theta_0 \frac{J_1(k\rho'\sin\theta_0)}{\rho'} \quad , \tag{10}$$

which shows the desired result: it takes an initial amplitude dependence of $\cos\theta$ to produce the familiar Airy-disk diffraction pattern. The standard problem of a constant-amplitude wave, when done with the precision used here, does not have a closed-form solution.

## 4. The Relation Between Before-Focus and After-Focus Wave Fronts

We now put $U(\rho', 0)$ from Eq. (10) in place of $U(x', y', 0)$ in the second line of Eq. (5) to find the amplitude on the far side of focus for the process shown in Fig. 1:

$$U(x, y, z) =$$
$$-i\cos\theta \frac{\exp(ikr)}{r} \iint_{x',y'} -iU_0 \frac{\sin\theta_0}{\lambda} \frac{J_1(2\pi\rho' \sin\theta_0 / \lambda)}{\rho'} \exp\left[-i2\pi\left(\frac{xx' + yy'}{\lambda r}\right)\right] dx' dy' \tag{11}$$

$$= -U_0 \cos\theta R\left(\frac{\sqrt{x^2 + y^2}/\lambda r}{\sin\theta_0 / \lambda}\right) \frac{\exp(ikr)}{r} = -U_0 \cos\theta R\left(\frac{\theta}{\theta_0}\right) \frac{\exp(ikr)}{r} \quad ,$$

where the integral, being the 2D Fourier transform of a circularly symmetric function, is taken from the first entry in Table 12.2 on p. 249 of Bracewell[5], with $R(u)$ used in place of Bracewell's $\Pi(u)$ (which $= 0$ for $|u| > \frac{1}{2}$). The relations $\sin\theta = (x^2 + y^2)^{1/2}/r$ and $R(\sin\theta/\sin\theta_0) = R(\theta/\theta_0)$ have also been used.

Thus we see that an initial wave front given by $U_0 \cos\theta R(\theta/\theta_0)\exp(-ikr)/r$, after passing through focus, becomes $-U_0\cos\theta R(\theta/\theta_0)\exp(ikr)/r$. The overall minus sign is the product of two factors of $-i$, each one of which denotes a phase loss of one-quarter cycle. Thus, compared to a plane wave traveling the distance from $z_-$ to $z$, the wave passing through focus has lost one-half cycle. This is the Gouy phase shift. A detailed description of the Gouy shift lies outside Fraunhofer diffraction (see, for example, p. 448 of Born and Wolf[1]). Observe that the Gouy shift is independent of the inclination factor, *i.e.*, of whether the RS or FK diffraction integral is used.

The relation between the before-focus wave and the after-focus wave is easy to generalize from the specific case considered above because the after-focus wave is found by twice applying a forward Fourier transform to the before-focus wave. The reader is reminded that the double forward transform of $f(x)$ is $f(-x)$, so, using the function $g(u, v)$ of Sec. 3, passing through focus changes $g(u, v)$ to $g(-u, -v)$ or $f[\theta(u, v), \phi(u, v)]$ to $f[\theta(-u, -v), \phi(-u, -v)]$. Now, $\sin\theta = (u^2 + v^2)^{1/2}$ and $\tan\phi = v/u$, so only $\phi$ is affected by sign changes of $u$ and $v$. Inverting the tangent function requires taking the signs of $u$ and $v$ into account: if $\tan^{-1}(v/u) = \phi$ then $\tan^{-1}[(-v)/(-u)] = \phi + \pi$. Thus, if Eq. (11) were reproduced for an arbitrary angular dependence of the original converging wave, *i.e.*, for $f(\theta, \phi)$ instead of $f(\theta) = \cos\theta R(\theta/\theta_0)$, the result would be $-U_0 f(\theta, \phi+\pi)\exp(ikr)/r$. Physically, this says that the wave front after focus is the same as the wave front before focus, but rotated 180° about the *z* axis, or, equivalently, inverted through the origin of the (*x*, *y*) plane, and with the Gouy phase shift applied.

## 5. Conclusion

By using the RS diffraction integral, keeping *r* (instead of approximating it) in Eqs. (5) and (6), and carefully considering the approximations that are made, this paper presents a complete description of Fraunhofer diffraction, in which *only* the Fraunhofer approximation has been made. The principal difference between the results presented here, given in Eqs. (5) and (8), and those of conventional treatments is that they contain the inclination factor, $\cos\theta$. In the overwhelming majority of optics problems, this factor can be safely approximated by 1, but Lucke[3] shows that describing Poisson's spot is an outstanding counterexample, and it is pointed out in Sec. 3 of this paper that $\cos\theta \approx 1$ becomes problematic in the case of a focus formed at a low *f*-number. Also, Eq. (8) is a Fourier transform from direction cosines to Cartesian coordinates, a form that is familiar to radio astronomers[4] (though it appears there in a different imaging context), but not to most optics workers.

The exact nature of the Fourier transform relation between the source and calculated wave fronts has been demonstrated for both the forward and inverse Fraunhofer problems, and the two problems combined to shown how the Gouy phase shift appears in Fraunhofer diffraction. It has been shown that a constant-amplitude converging wave front passing through a circular aperture does not produce exactly the well-known Airy pattern, though the difference is small for the values of $\theta$ normally encountered in optical problems.

## Acknowledgement

This work was partially supported by the National Aeronautics and Space Administration through the California Institute of Technology's Jet Propulsion Laboratory.